\documentclass[aps,showpacs,twocolumn]{revtex4-1}
\usepackage[pdftex,plainpages=false,colorlinks=true,linkcolor=blue, citecolor=blue, urlcolor=blue]{hyperref}

\usepackage{amssymb}
\usepackage{epsfig}
\usepackage{graphicx}
\usepackage{amsmath}
\usepackage{array,color}
\usepackage{natbib}

\begin{document}

\title{d-wave superconductivity in the frustrated two-dimensional periodic Anderson model}

\author{Wei Wu$^{1}$ and A.-M.-S. Tremblay$^{1,2}$}
\affiliation{
$^1$D{\'e}partement de Physique and RQMP, Universit{\'e} de Sherbrooke, Sherbrooke, Qu{\'e}bec, Canada \\
$^2$Canadian Institute for Advanced Research, Toronto, Ontario, Canada
}

\date{\today}

\begin{abstract}
Superconductivity in heavy-fermion materials can sometimes appear in the incoherent regime and in proximity to an antiferromagnetic quantum critical point. Here we study these phenomena using large scale determinant quantum Monte Carlo simulations and the dynamical cluster approximation with various impurity solvers for the periodic Anderson model with frustrated hybridization. We obtain solid evidence for a $d_{x^2-y^2}$ superconducting phase arising from an incoherent normal state in the vicinity of an antiferromagnetic quantum critical point. There is a coexistence region and the width of the superconducting dome increases with frustration. Through a study of the pairing dynamics we find that the retarded spin-fluctuations give the main contribution to the pairing glue. These results are relevant for unconventional superconductivity in the Ce-$115$ family of heavy-fermions.
\end{abstract}


\maketitle

\section{Introduction}

d-wave superconductivity in proximity to a quantum critical point has been found in many compounds, such as layered and quasi one-dimensional organic superconductors, iron pnictides, cuprate superconductors and heavy fermion systems. 
In particular, many examples of quantum critical points (QCPs) in  heavy fermion materials have been found, making them an important testing ground for theories of quantum criticality in relation to superconductivity~\cite{PColeman2007review}. This relationship is the main problem that we consider.

Heavy fermion behavior arises when localized $f$-electron bands are hybridized with conduction-electron bands. This hybridization leads to the RKKY interaction between $f$-electron local moments, but also to the eventual screening of the local moments by the Kondo effect. These competing tendencies are summarized by Doniach's phase diagram~\cite{Doniach1977a, *[{For reviews, see }]Hewsonbook,*PColeman2007review} where antiferromagnetic and Fermi-liquid phases both appear. The zero-temperature transition between these phases is generally believed to be a quantum critical point (QCP). Remarkably, superconductivity appears in the vicinity of this QCP~\cite{steglich2013routes, *[{For a nontrivial example of \textit{Kondo-breakdown} QCP induced superconductivity, see  }] park2006,*shishido2005}, even though the nature of the QCP can change depending on the Kondo deconstruction energy scale $E^*_{loc}$~\cite{FGrosche2000, Moriya1973, *[{For a review of the Hertz-Millis-Moriya theory, see }]Lohneysen2007}~\cite{schroder1998,*schroder2000onset,Coleman1999,*si2001locally, *Paul2007, *gegenwart2008,*si2010heavy,*si2013kondo,sun2005}.
 

A few heavy-fermion superconductors~\cite{Pfleiderer2009},  for example PuCoGa$_5$ and CeCoIn$_5$, show the peculiarity~\cite{PColeman2007review} that $d_{x^2-y^2}$ pairing develops out of an incoherent metallic state~\cite{KIzawa2001,*allan2013, *Bzhou2013,*sidorov2002, daghero2012}.  CeCoIn$_5$ belongs to the quasi-2D Ce-$115$ materials that have an easily accessible transition temperature ($\sim 2 K$) below which superconducting and magnetic properties can be precisely measured 
~\cite{Hegger2000,*Petrovic2001,*Petrovic2001Ir, *curro2005}. They are especially interesting because the itinerant-to-localized transition of $4f$ electrons can be readily obtained by applying pressure 
or changing the chemical composition~\cite{Haule2010, LSarrao2007review}. Moreover, the observation of  an evolving superconducting dome in the vicinity of the magnetic QCP strongly suggests that they are candidates for antiferromagnetic (AFM) spin-fluctuation mediated superconductivity~\cite{Beal-Monod:1986,*Scalapino:1986,*Miyake:1986}\cite{THu2012, JVDyke2014}. The proximity of antiferromagnetism to d-wave superconductivity is also observed in many other strongly correlated systems, such as cuprate superconductors and layered organic superconductors.

In some unusual cases ~\cite{yuan2003,Nicklas2004,LShu2011,Griveau2005,*savrasov2006},
heavy-fermion superconductors have been found in the absence of an obvious nearby magnetic QCP. The americium metal under high-pressure~\cite{Griveau2005}, the ``SCII'' phase of CeCu$_2$Si$_2$~\cite{yuan2003, Pourovskii2014} and heavily Yb doped CeCoIn$_5$~\cite{LShu2011} are examples. In an attempt to understand these superconductors, alternative scenarios have been proposed, such as the valence fluctuation hypothesis~\cite{miyake1999superconductivity,Bauer2012}, or the composite pairing theory \cite{Flint2008, Flint2010, OErten2014}. Yet, up to now, no concensus has been reached.
There are also materials where, despite the absence of an obvious QCP, spin-fluctuation mediated pairing is considered essential, for instance  CeIrIn$_5$~\cite{Kasahara2008, Shang2014, yang2014scaling}. 

On the theoretical side, the two-dimensional periodic Anderson model (PAM), the Kondo-lattice model (KLM) or
the degenerate Coqblin-Schrieffer model ~\cite{Rice1985} are expected to capture the essential physics of spin-fluctuation mediated superconductivity in Ce-$115$ compounds. 
Previous analytical studies include large-$N$ approaches~\cite{PColeman2007review}, mean-field theory ~\cite{YLiu2012} and phenomenological models of fermions coupled to fluctuating Bose modes~\cite{nishiyama2013}.
The PAM/KLM models have also been treated with the variational method~\cite{asadzadeh2014}, exact diagonalization (ED) and  DMRG  calculations on small clusters~\cite{Xavier2008}. In these studies, Heisenberg exchange 
is usually artificially added to simulate the RKKY interaction. 
 
Here we show, using large-scale determinant quantum Monte Carlo (DQMC) simulations~\cite{bss1981} as well as the dynamical cluster approximation (DCA)~\cite{ThMaier2005review} \footnote{Further details on the methodology can be found in Appendix B.} that 
$d_{x^2-y^2}$ superconductivity can arise out of an {\it incoherent} metallic phase in the frustrated PAM. 
Heisenberg exchange is not artificially added, it arises naturally from the PAM. We demonstrate that the width of the superconducting dome surrounding the QCP can be increased by increasing frustration. Based on the magnetic susceptibility and the anomalous self-energy, we find that the driving force for pairing
in this model comes primarily from  
retarded antiferromagnetic spin fluctuations. This reinforces the hypothesis that this mechanism applies to Ce-$115$.

This paper is organized as follows. In Sec. II we introduce the frustrated periodic Anderson model. Evidence for $d_{x^2-y^2}$ pairing is first presented in Sec. III using DQMC and DCA calculations done with a quantum Monte Carlo impurity solver (CTQMC)~\cite{Rubtsov2005}. DCA results presented in the rest of Sec. III allow us to discuss successively: quasi-particle coherence, the relation between the antiferromagnetic QCP and superconductivity, and finally the origin of pairing. The discussion in Sec. IV also contains material-specific comments. We conclude in Sec. V. Our model is justified in more details in Appendix A and additional information on the DCA method can be found in Appendix B. 

\section{Periodic Anderson Model with Frustrated Hybridization}

Frustration, Kondo coupling strength and $f$-orbital degeneracy 
determine Doniach's phase diagram of heavy-fermion systems~\cite{si2013kondo}. These effects are embodied in the frustrated periodic Anderson model on a two-dimensional square lattice~\cite{WeberVojta:2008}, with Hamiltonian
\begin{equation}
\begin{split}
H = \sum_{k,\sigma}\epsilon_{k}c_{k,\sigma}^{\dagger}c_{k,\sigma}+\sum_{k,\sigma}\epsilon^{f}f_{k,\sigma}^{\dagger}f_{k,\sigma}\\
+\sum_{k,\sigma}V_{k}(f_{k,\sigma}^{\dagger}c_{k,\sigma}+\text{h.c.})
+\sum_{i }U(n_{f}^{\uparrow}-\frac{1}{2})(n_{f}^{\downarrow}-\frac{1}{2} )
\end{split} 
\end{equation}
where {${k,\sigma,i}$} are the momentum, spin and lattice site indices respectively.
$n^{\sigma}_f$ denotes the occupation number operator for $f$-orbitals. The conduction band dispersion relation 
is chosen as $\epsilon_{k}=-2t[\cos(k_x)+\cos(k_y)]$. The nearest-neigbour 
hopping integral $t$ is taken as the energy unit throughout this paper.  We neglect the dispersion of $f$-orbitals and $f$-orbital degeneracy. The $f$ energy level $\epsilon^{f}$ is set to zero and the fillings are $\langle n_f \rangle \sim 1$, $\langle n_c\rangle \sim 0.9$. The strength of the Kondo coupling and of the RKKY interaction is determined by the combined effects of the $f-c$ hybridization $V_{k} = V+2V'[\cos(k_x)+\cos(k_y)]$ and of $U$, the screened Coulomb repulsion between $f$ electrons. In an antiferromagnetic configuration of the conduction electrons, the on-site hybridization $V$ and the hybridization with nearest-neighbors $V'$ lead to competing effective interactions with the $f$ electron. The frustration is maximal when $V$ and $V'$ are of the same order of magnitude, as in Ce-$115$~\cite{allan2013,yang_emergent_2012}. Further discussion of the model appears in Appendix A. 

\begin{figure}[t]
\includegraphics[bb=0bp 0bp 600bp 440bp,scale=0.4]{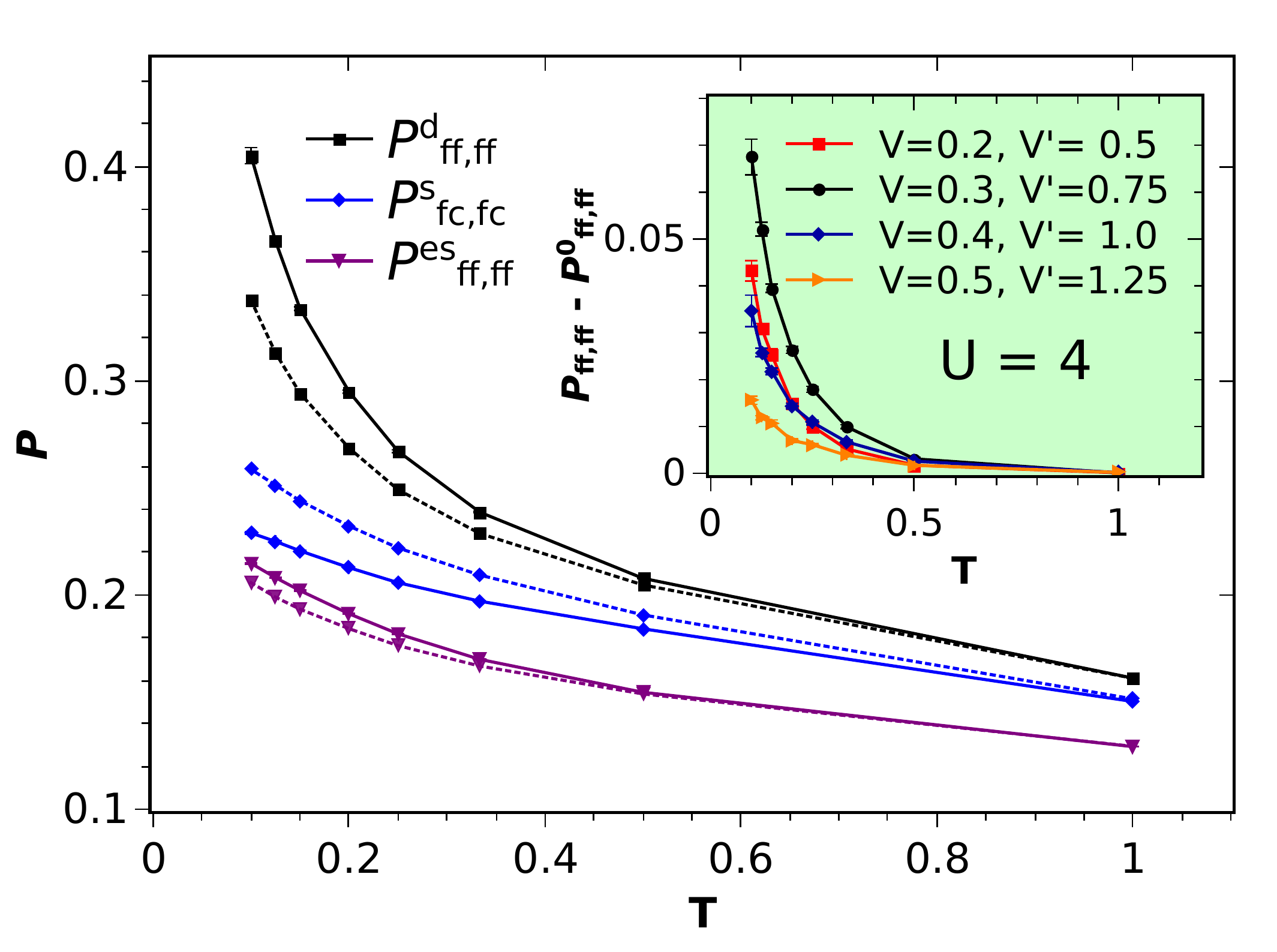}
\caption{(color online) Pairing susceptibilities for $V=0.3, V'=0.75, U=4$ calculated with DQMC on a $12\times12\times2$ lattice with periodic boundary
conditions. The Trotter decomposition imaginary time interval is
 $\Delta \tau=0.0625$.
Solid curves denote susceptibilities $P$ including vertex contributions while dashed 
lines represent the bubble contribution $P^0$. The inset shows
the effective pairing interaction $P-P^0$ for $d_{x^2-y^2}$ pairing in the $(ff,ff)$ channel, keeping the ratio $V'/V$ constant but changing the hybridization gradually from $V=0.2,V'=0.5$
to $V=0.5,V'=1.25$. Increasing the hybridization makes the $f$-electrons more itinerant.
}
\end{figure}

\begin{figure*}[t]
\includegraphics[bb=4bp 13bp 537bp 235bp,scale=0.75]{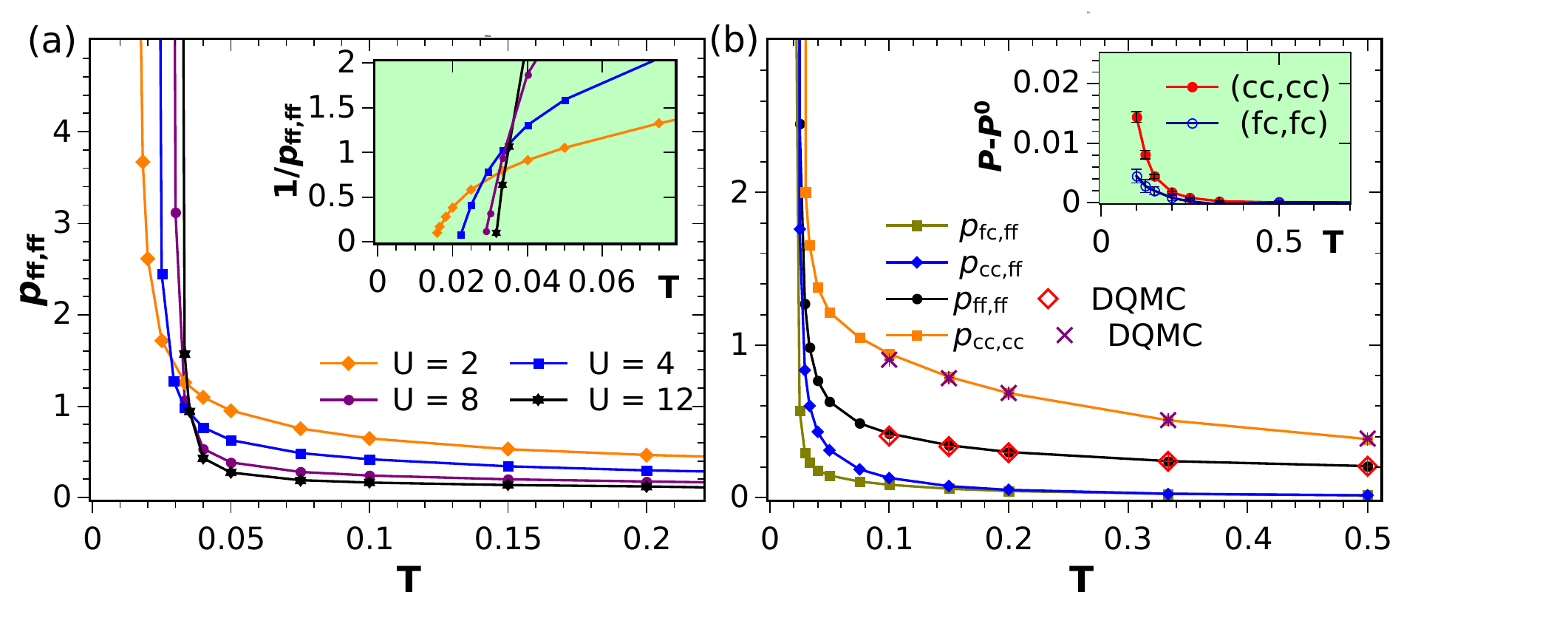}
\caption{(color online) $d_{x^2-y^2}$  pairing susceptibility as a function of temperature from DCA calculations on $2\times 2\times2$ clusters. (a) ($ff,ff$) channel for $U$= 2, 4, 8 , 12 at constant $V^2/U$ and $V'=2.5V$. The inset shows that the inverse pairing susceptibilities extrapolate to zero, signaling that $T_c$ increases with $U$. (b) Comparison of different channels for $V=0.3$, $V'=0.75$ and $U=4$.
Available DQMC results at large $T$, shown by crosses and open diamonds, are in excellent agreement with those of DCA. The inset shows the effective pairing interaction in the $(cc,cc)$ and  $(fc,fc)$ channels calculated by DQMC. They are small 
compared to their $(ff,ff)$ counterpart shown in the inset of Fig.~1. 
}
\end{figure*}

\section{Results}

After we present evidence for $d_{x^2-y^2}$ pairing, we discuss the question of quasi-particle coherence then the phase diagram and conclude with the origin of pairing.
 
\subsection{$d_{x^2-y^2}$ pairing}
To reveal 
the many-body correlation effects on superconductivity in an unbiased way, 
we show the result of DQMC calculations of the pairing susceptibility $P$, defined by 
\begin{equation}
\begin{split}
P_{\alpha\beta,\gamma\delta} =  \frac{1}{N}\times\frac{1}{G}\sum_{i,j}\sum_{r,r'}g(r')g^{*}(r)\\
\int_{0}^{\beta}\left\langle d_{\alpha,j+r',\downarrow}(\tau)d_{\beta,j,\uparrow}(\tau)d_{\gamma,i,\uparrow}^{\dagger}(0)d_{\delta,i+r,\downarrow}^{\dagger}(0)\right\rangle d\tau
\end{split}
\end{equation}
where the Greek indices represent either conduction $c$ or localized $f$ electron operators, $g(r)$ is the form factor in real space and $G=\sum_{r}|g(r)|^2$ is the normalization factor. Let $P^{0}$ be the bubble contribution without vertex corrections. For a given pairing channel, the sign of $P-P^0$ basically reflects 
whether the pairing is favoured (positive) or not (negative)~\cite{scalapino2006book}. 

Fig.~1 displays our DQMC results for $d_{x^2-y^2}$, $s$,  and extended $s$  wave pairing susceptibilities \footnote{The form factors for $d_{x^2-y^2}$, $s$  and extended $s$ wave in $k$ space are, respectively, $cos(kx)-cos(ky)$, 1 and $cos(kx)+cos(ky)$.} as a function of temperature $T$ for  $V=0.3$, $V'=0.75$ and $U=4$.
We learn that among those various pairing channels, $d_{x^2-y^2}$ dominates~\cite{allan2013} since the effective pairing interaction $P-P^0$ increases rapidly as $T$ is lowered (inset of Fig.~1), suggesting that a divergent susceptibility would occur at the Berezinsky - Kosterlitz - Thouless (BKT) transition temperature $T_{BKT}$~\cite{ThMaier2005}. This contrasts with local $s$-wave pairing where the effective interaction $P-P^0$ decreases, becoming more negative as $T$ is lowered. This rules out $s$-wave pairing \footnote{Such pairing is seen for the KLM~\cite{Bosensiek2013} that does not have the nearest-neigbour hybridization.}.

By varying $V$ and $V'$, the itinerant character of $f$-electrons can be adjusted. This is shown in the inset of Fig.~1, where DQMC results suggest that, at fixed $V'/V$, the pairing strength has a maximum for intermediate $V$ and $V'$, namely for $V$ = 0.3 and $V'$=0.75. This occurs
in the vicinity of an antiferromagnetic phase transition, as we shall see.

Further insight into the nature of pairing and on the phase diagram is provided by DCA~\cite{Hettler:1998,ThMaier2005review}, which allows us to reach much lower temperature than DQMC.
Following Ref.~\onlinecite{Lin2012}, the pairing susceptibilities are obtained from small pinning fields in the linear response region.
The results are shown in Fig.~2 for temperatures as low as $T\sim 0.015$. The inset of Fig.~2a shows that
$T_c$ can be extrapolated from the diverging susceptibilities.
Since in Ce$-115$ materials the Coulomb repulsion $U$ of $4f$ orbitals is expected to be significantly larger than
the LDA bandwidth, ($U \sim 5$~eV vs $W \sim 400$~meV)~\cite{Costa-Quintana2003}, we study the evolution of pairing upon approaching the extended Kondo limit of the PAM model. In Fig.~2a we display the $d_{x^2-y^2}$ pairing susceptibility in the ($ff,ff$)
channel, keeping $V^2/U$ constant at $V^2/U \approx 0.0225$, and increasing $U$~\cite{RDong2013}. The behavior as a function of $T$ clearly suggests that $d_{x^2-y^2}$ pairing is stable in the extended Kondo limit. In fact, the estimated $T_c$ grows with increasing $U$, despite the fact that at large $T$ 
the susceptibility is suppressed with increasing $U$. At large $T$, where vertex corrections are not important, $U$ reduces the low-energy density of states (DOS), decreasing the pairing susceptibility.
By contrast, at low $T$, increasing $U$ drastically diminishes the higher order frustrated exchange terms, 
therefore enhancing the magnetic interaction vertices that mediate Cooper pairing.

Since a heavy fermion is composed of
both $f$ and $c$ electrons,  
$a_{k\sigma}^{\dagger}  =  u_{k}c_{k\sigma}^{\dagger}+v_{k}f_{k\sigma}^{\dagger}$,
the heavy fermion Cooper pair $\langle a^{\dagger}_{k,\uparrow} a^{\dagger}_{-k,\downarrow} \rangle$ has 
four different $d_{x^2-y^2}$ susceptibilities that should diverge simultaneously at $T_c$. This is shown in Fig.~2b. Since the RKKY coupling between the local moments of the $f$- orbitals is emergent in the PAM, the magnetic ``pairing glue'' originates from the $f$-orbitals. This is consistent with Fig.~2b. Note that at large $T$, when
Kondo screening is weak, the $(fc, ff)$ and $(cc,ff)$ 
channels are strongly suppressed and the effective pairings $P-P^0$ in the $(cc,cc)$ and $(fc,fc)$ channels, shown in the inset, are small compared with the $(ff,ff)$ channel in the inset of Fig.~1.  This differs from the prediction of the composite pairing theory~\cite{Flint2008} for a two-channel Kondo lattice model. That model has a different source of pairing, leading to a dominant $fc$ pairing channel. 

 \begin{figure*}
 	\includegraphics[bb=0bp 0bp 510bp 330bp,scale=0.7]{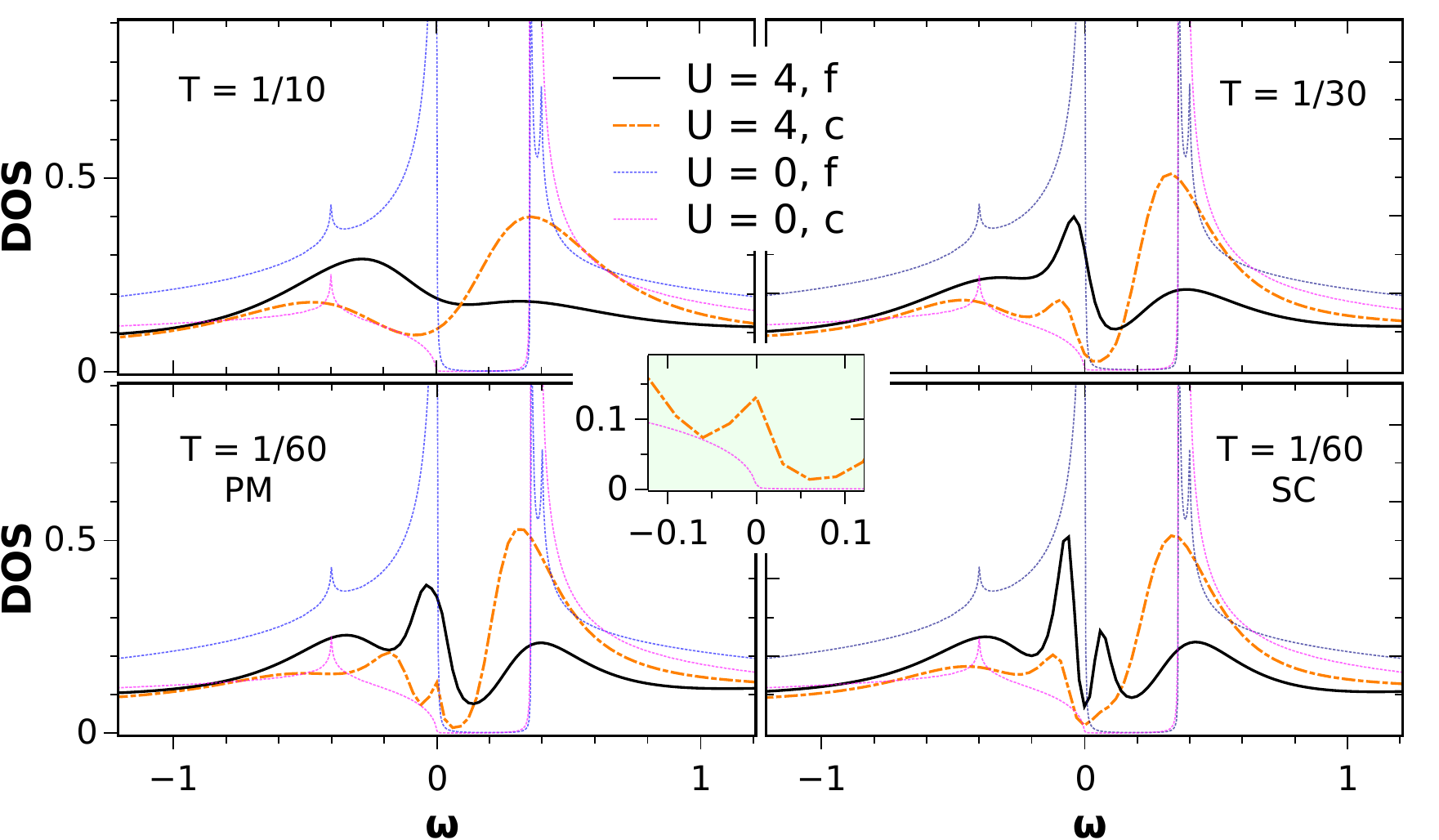}
 	\caption{(color online) Local density of states of $f$ and conduction bands at $V=0.4$, $V'=0.8$ and $U=4$ for various temperatures.
 		For reference, the DOS for the non-interacting case ($U=0$) is repeated on all panels. The bottom-left panel shows the DOS when superconductivity is artificially removed.  The corresponding low-energy conduction band DOS in the underlying normal state is blown-up in the inset.
 	}
 \end{figure*}
 
\subsection{Quasi-particle Coherence}
We stress that complete screening of the local moments is not a necessary condition to find diverging 
pairing susceptibilities. In fact, we find that complete screening of $f$ moments does not occur
even when the QCP is approached. This is confirmed by the fact that the magnetic susceptibility increases with decreasing $T$ (not shown), meaning that in our case the Fermi surface is small at the antiferromagnetic to paramagnetic
transition.

This is further illustrated by the DOS obtained from maximum entropy analytic continuation~\cite{jarrell1996bayesian} of the local Matsubara Green functions in Fig.~3. The non-interacting result, shown in light colors, is repeated for reference on all panels. As shown on the top-left panel of Fig.~3, at large $T$ the heavy-fermion quasiparticle is absent, and because of the intense scattering by $f$ local moments, the effective hybridization between the conduction band and the $f$-band is reduced, leading to a strong suppression of the hybridization gap, and to a larger DOS for the conduction band  at the Fermi level (red line). This can be understood in the scenario of $f$-orbital selective Mott transition (OMST)\cite{LDeLeo2008, MVojta2010}. The \textit{d}- wave superconducting gap that appears on the emergent quasiparticle peak at low $T$ is displayed on the lower-right panel.  On the bottom-left panel the superconducting order parameter is suppressed. One then sees that in the underlying normal state, the DOS of the conduction band develops a peak at the Fermi level, reflecting the increased damping of the low-energy quasiparticles on the corresponding $f$-orbital. 



\begin{figure*}
\includegraphics[bb=0bp 25bp 464bp 367bp,scale=0.6]{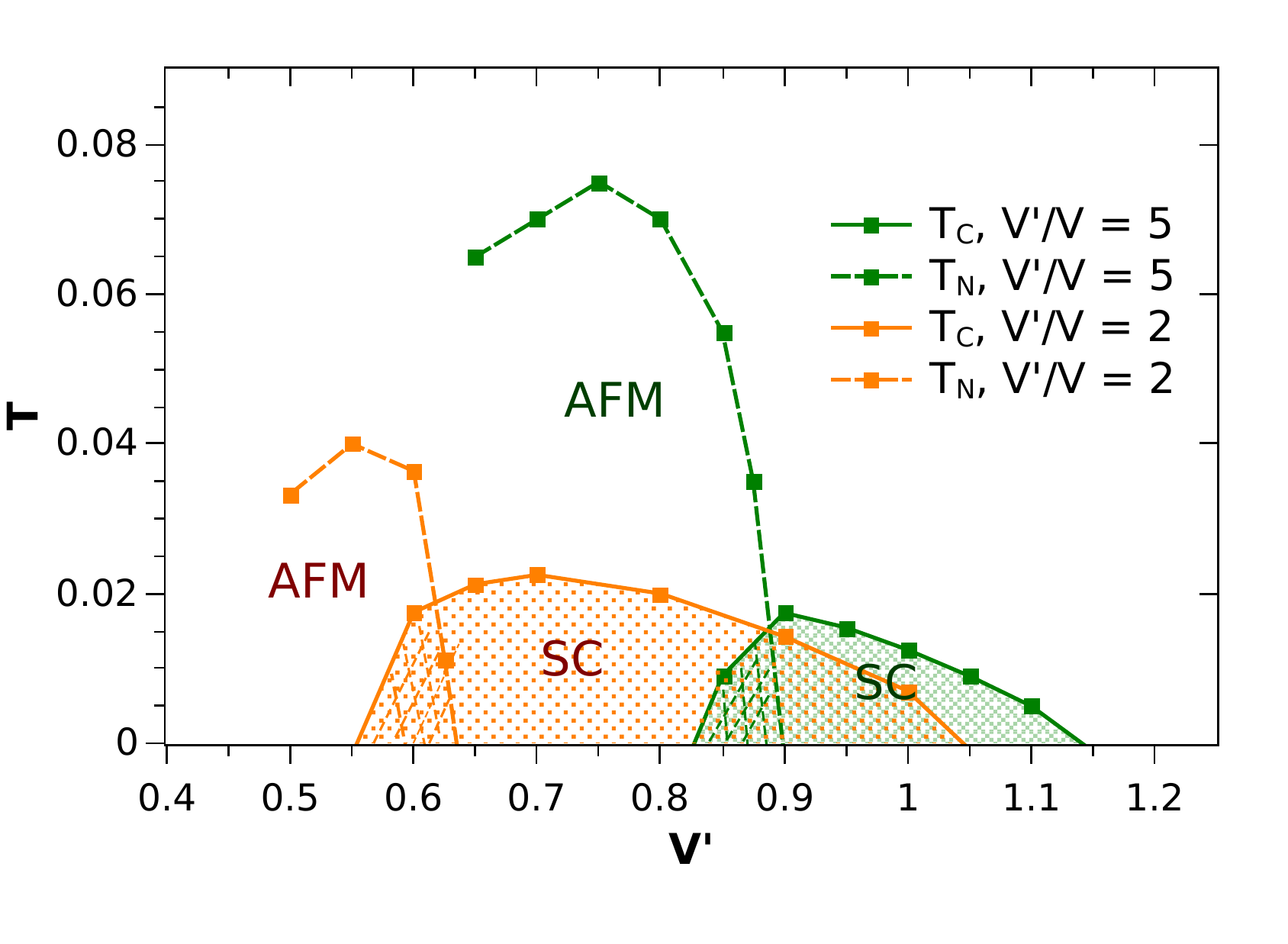}
\caption{(color online) Phase diagram of the frustrated 2D PAM model within $2\times2\times2$ DCA at $U=4$.
Two groups of result with different $V'/V$ ratios are shown. Darker lines (green) are drawn for the more frustrated case ($V'/V=2$),
whereas lighter lines (orange) are for the less frustrated case ($V'/V=5$). The N\'eel temperature $T_N$ (dashed lines) and the superconducting $T_c$ (full lines) are drawn where the respective order parameters vanish. For both values of $V'/V$ there is a uniform coexistence region indicated by crosshatch. 
}
\end{figure*}

\subsection{Phase diagram, QCP and frustration} \label{SectionPD, QCP,Frustration}

To mimic the Doniach phase diagram, where the ratio of Kondo to RKKY couplings is the control parameter, we plot in Fig.~4 the phase boundaries in the $T-V'$ plane for two fixed values of the frustration, $V'/V=2$ and $V'/V=5$.  In the limit of small hybridization, the Kondo screening and antiferromagnetic RKKY correlations in the PAM/KLM both vanish, leaving only local moment fluctuations in the orbital-selective Mott insulator with large entropy at non-zero temperatures~\cite{LDeLeo2008}. As $V'$ increases, the Kondo energy scale and RKKY interaction both increase. First, RKKY dominates over Kondo screening and the antiferromagnetic ground state appears in coexistence with superconductivity.  
Increasing $V'$ again drives the system across the QCP. Then, long-range RKKY correlations are gradually quenched by Kondo screening, and eventually superconductivity is destroyed when $f$ electrons become too itinerant. 

It is striking that the superconducting dome follows the change in the position of QCP with changing magnetic frustration ($V' /V$ ratio). This result vividly illustrates the intrinsic connection between the QCP and superconductivity in the PAM model.

\subsection{Retardation and origin of pairing}

The origin of pairing may be deduced from the dynamical processes entering the real part of the anomalous self-energy $I_{\Sigma}(\omega)$~\cite{TMaier2008dynamics} at zero frequency and from the cumulative order parameter $I_{G}(\omega)$~\cite{Bkyung2009pairing}, defined respectively by  
\begin{eqnarray}
I_{\Sigma}(\omega) = \frac{\frac{2}{\pi}\int^{\omega}_{0}\frac{\Sigma''_{a}(\omega')}{\omega'}dw'}
{\frac{2}{\pi}\int^{\infty}_{0}\frac{\Sigma''_{a}(\omega')}{\omega'}dw'},\quad & 
I_{G}(\omega) = \int^{\omega}_{0}\frac{d\omega'}{\pi}F''(\omega')
\end{eqnarray}
where $\Sigma''_{a}(\omega)$ is the imaginary part of the anomalous self-energy $\Sigma^{i,i+r}_{a}(\omega)$ while $F''(\omega)$ is the imaginary part of the retarded lattice Gork’ov function $F''(\omega) = -\mathrm{Im} \int^{\beta}_{0} d\tau \langle c_{i+r}(\tau) c_{i}(0) \rangle e^{i(\omega+i\eta) \tau}$, 
with $i$ and $i+r$ nearest neighbours.

These functions are plotted in Fig.5 along with the imaginary part of the antiferromagnetic spin susceptibility $\chi''(q=(\pi,\pi),\omega)$. As in the case of cuprates~\cite{TMaier2008dynamics,Bkyung2009pairing}, we find the that the Cooper pairs initially form over an energy range comparable to that over which antiferromagnetic fluctuations develop. The dependence of pairing on the RKKY interaction strength can clearly be seen in Fig.~5,
where increasing $U$, hence decreasing the RKKY interaction, shifts the peaks of both $I_{\Sigma}(\omega)$ and $\chi''(q=(\pi,\pi),\omega)$ towards the low-energy side. Frustrating magnetism in the conduction band by adding a next-nearest neighbour hopping $t'$ also leads to the same correlation between the two quantities: the characteristic frequency of the spin fluctuations decreases along with characteristic frequencies in both $I_{\Sigma}(\omega)$ and $I_{G}(\omega)$. Our results thus confirm that the retarded spin-fluctuations mediate \textit{d}-wave superconductivity in heavy fermion superconductors.

In addition to the contribution of low-frequency (retarded) spin-fluctuations, Fig.5b shows that there can be a significant gain in pairing for an energy scale set by the upper Hubbard band of the $f$ electrons. This high-frequency (more instantaneous) contribution to pairing is much larger than what is found in the case of cuprates~\cite{Bkyung2009pairing}. This is probably because the upper Hubbard band seen from the $f$ electron point of view is still in the conduction band. In other words, $I_{G}(\omega)$ is enhanced by the large RKKY interaction~\cite{TMaier2008dynamics} that results from intermediate $U$ and moderate conduction band frustration. In realistic Ce$-115$ materials the RKKY interaction is believed to be small~\cite{LSarrao2007review}, hence the high-frequency contribution to pairing should be less important.
 
 \begin{figure*}
\includegraphics[bb=0bp 20bp 320bp 270bp,scale=0.9]{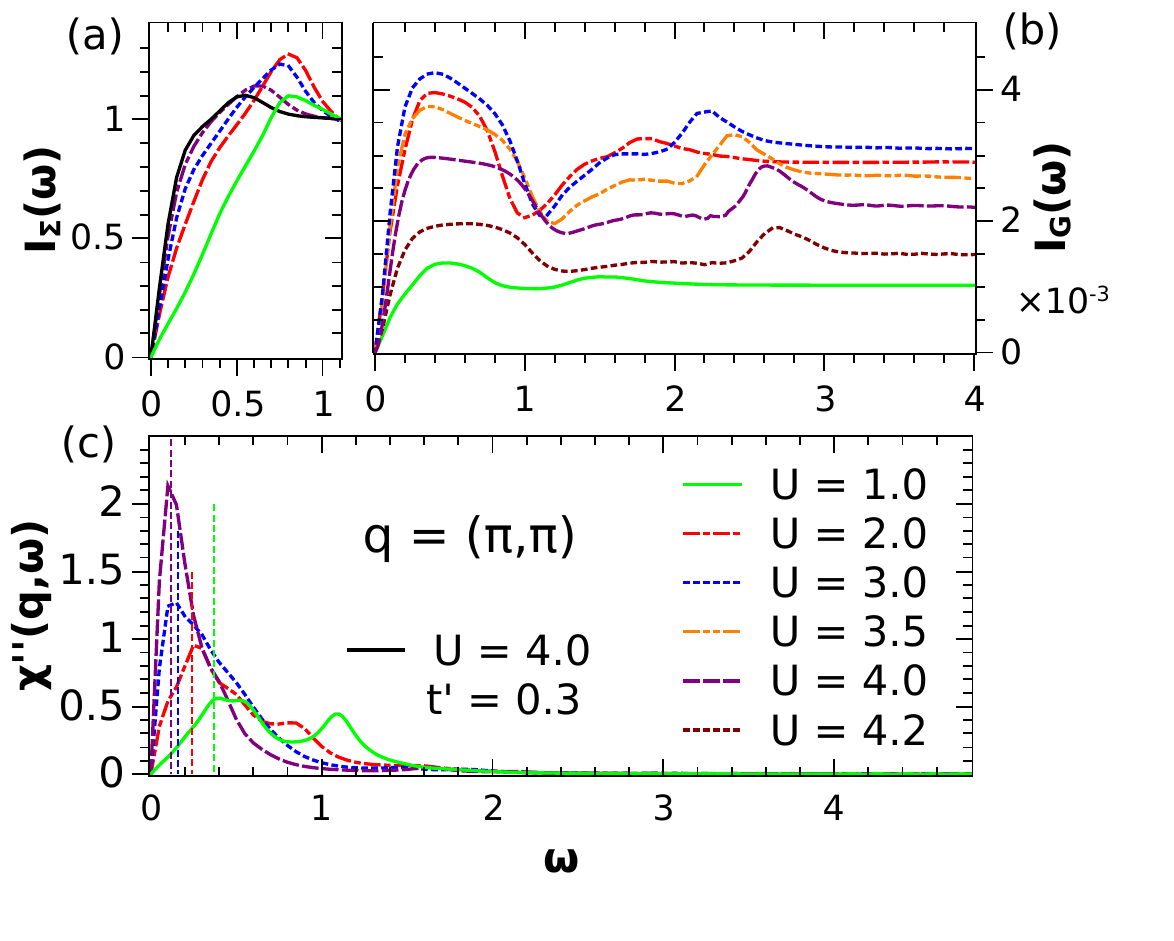}
\caption{(color online) (a) The low energy part of $I_{\Sigma}(\omega)$, (b) $I_{G}(\omega)$ and (c) the imaginary part of the antiferromagnetic
susceptibility $\chi_{ff,ff}''(q, \omega)$ as a function of real frequency $\omega$ for various $U$. Results are obtained with $2\times2\times2$ DCA, using a Lanczos impurity solver. The broadening factor is chosen as $\eta=0.125$. Vertical dashed lines in (c) show energies where $I_{\Sigma}(\omega)$ reaches half of its maximum value for $U=1,2,3,4$ at fixed frustration and Kondo coupling: $V=0.3$ and $V'=0.6$.
}
\end{figure*}

\section{Discussion}

Spin-fluctuation mediated pairing theory finds solid support in both our DQMC and DCA results. As noted in Sec.~\ref{SectionPD, QCP,Frustration}, this is manifested clearly by the correlation between the location of the superconducting dome and that of the antiferromagnetic QCP when magnetic frustration ($V'/V$ ratio) is varied. The analysis of the frequency-dependent pairing correlations based on a Lanczos exact-diagonalization solver further confirms the spin-fluctuation mediated origin of superconductivity in the frustrated PAM. 

Our model has a Fermi surface resembling the $\alpha$ sheet of Ce$-115$ materials~\cite{aynajian2012}, where most of the pairing occurs. In the phase diagrams, illustrated in  Fig.~4, there is a coexistence region between antiferromagnetism and superconductivity, as observed in the Ce$-115$ family. We also note that the right-hand side of the superconducting dome does not move towards much larger $V'$ even when the QCP does. This can be understood as indicating that, in this region, the $f$ electrons become more itinerant, leading to a suppression of  antiferromagnetic fluctuations, even when there is a nearby QCP.

A few material-specific comments arise naturally. By comparing Fig.~3 with Fig.~1c of Ref.~\onlinecite{zhou_visualizing_2013}, one can verify that our model produces features in the low-energy density of states that are similar to those of CeCoIn$_5$.  The ratio of the maximum N\'eel temperature and the maximum superconducting $T_c$ for $V'/V=2$ in Fig.~4 is similar to that observed in CeRhIn$_5$~\cite{YashimaCeRhIn:2007}. Finally, to explain the appearance of a superconducting phase in CeIrIn$_{5}$, despite the absence of a nearby magnetic QCP, we note that the superconducting dome widens in parameter space when the system is more frustrated ($V'/V$ =2) so that, in this compound, the QCP might not be observable for physically accessible parameters.  

Although we expect our results to be relevant for materials where antiferromagnetic fluctuations due to localized $f$ electrons are important, there could be heavy fermion superconductors where the spin-fluctuation scenario doesn't apply.  CeCu$_2$Si$_2$ under high ambient pressure may be an example. Indeed, recent \textit{ab initio} calculations~\cite{Pourovskii2014} suggest the existence of an orbital transition in this material, which could be responsible for the underlying ``SCII'' phase. In other words, when two low-lying crystal-field levels become degenerate upon increasing pressure, they can compete to screen the $f$ local moment, eventually causing composite pairing~\cite{Flint2008,Flint2010}. To investigate cooperation and competition between composite pairing and magnetic pairing, one would need to consider the two-channel Kondo lattice model~\cite{Jarrell1997} or the two-channel periodic Anderson model.

\section{Conclusion}
 
The 2D PAM model with frustration induced by the real-space structure of the hybridization $V,V'$ between $f$ and $c$ electrons exhibits many features of the Ce$-115$ materials. Our unbiased DQMC large-cluster simulations suggest that in this model, 
$d_{x^2-y^2}$ pair correlations increase rapidly at low $T$, and are enhanced when the anti-ferromagnetic QCP is approached. The DCA-CTQMC calculations further confirm the existence of a $d_{x^2-y^2}$ superconducting phase that appears out of an incoherent metallic phase. Pairing is strongest on the $f$ electrons. In the $T-V'$ plane, the superconducting phase has a dome shape that surrounds the QCP of the antiferromagnetic phase. Finally, through an analysis of the frequency dependence of pairing, we have shown that $d$-wave superconductivity in this model is mediated by retarded spin fluctuations.  

\acknowledgments
We would like to thank A. Georges, M. Aichhorn, S. Burdin, P. Coleman, C. Lacroix, J. Schmalian, C. P\'epin, L. Taillefer and M. Vojta for useful discussions. We are grateful to D. S\'en\'echal for comments on the manuscript. This work was supported by the Natural Sciences and Engineering Research Council of  Canada (NSERC), and by the Tier I Canada Research Chair Program (A.-M.S.T.). Simulations were performed on computers provided by CFI, MELS, Calcul Qu\'ebec and Compute Canada.

\appendix

\section{Hybridization gap and further justifications of the frustrated 2D periodic Anderson model}
The dispersion relation for the PAM model of in Eq.~(1) reads,
\begin{eqnarray}
\begin{split}E_{k}^{\pm} & =\epsilon_{k}^{c}/2\pm\sqrt{(\epsilon_{k}^{c}/2)^{2}+V_{k}^{2}}\\
\end{split}
\end{eqnarray}
where $\epsilon_{k}^{c}$ is the conduction band dispersion and $V_{k}$ is the $k$-dependent
hybridization between the $f$-band and the conduction band. To incorporate frustration, $V_{k}$
includes both on-site hybridization $V$ and nearest-neighbour hybridization $V'$. When $V$ is not too large
compared to $V'$, say $|V|<4|V'|$, as in this paper, the hybridization gap $\Delta$ 
\begin{eqnarray}
\Delta  =|\frac{4VV'}{1+4V'^{2}}|,
\end{eqnarray}
lies above (or below, depending on the 
sign of $V/V'$) the Fermi level.
Note that the superconducting and hybridization gaps, as seen in the low-energy density of states Fig.~3, are qualitatively similar to those of the experimental results displayed in Fig.~1c of Ref.~\onlinecite{zhou_visualizing_2013}.

\begin{figure}[!b]
\includegraphics[bb=0bp 15bp 520bp 435bp,scale=0.4]{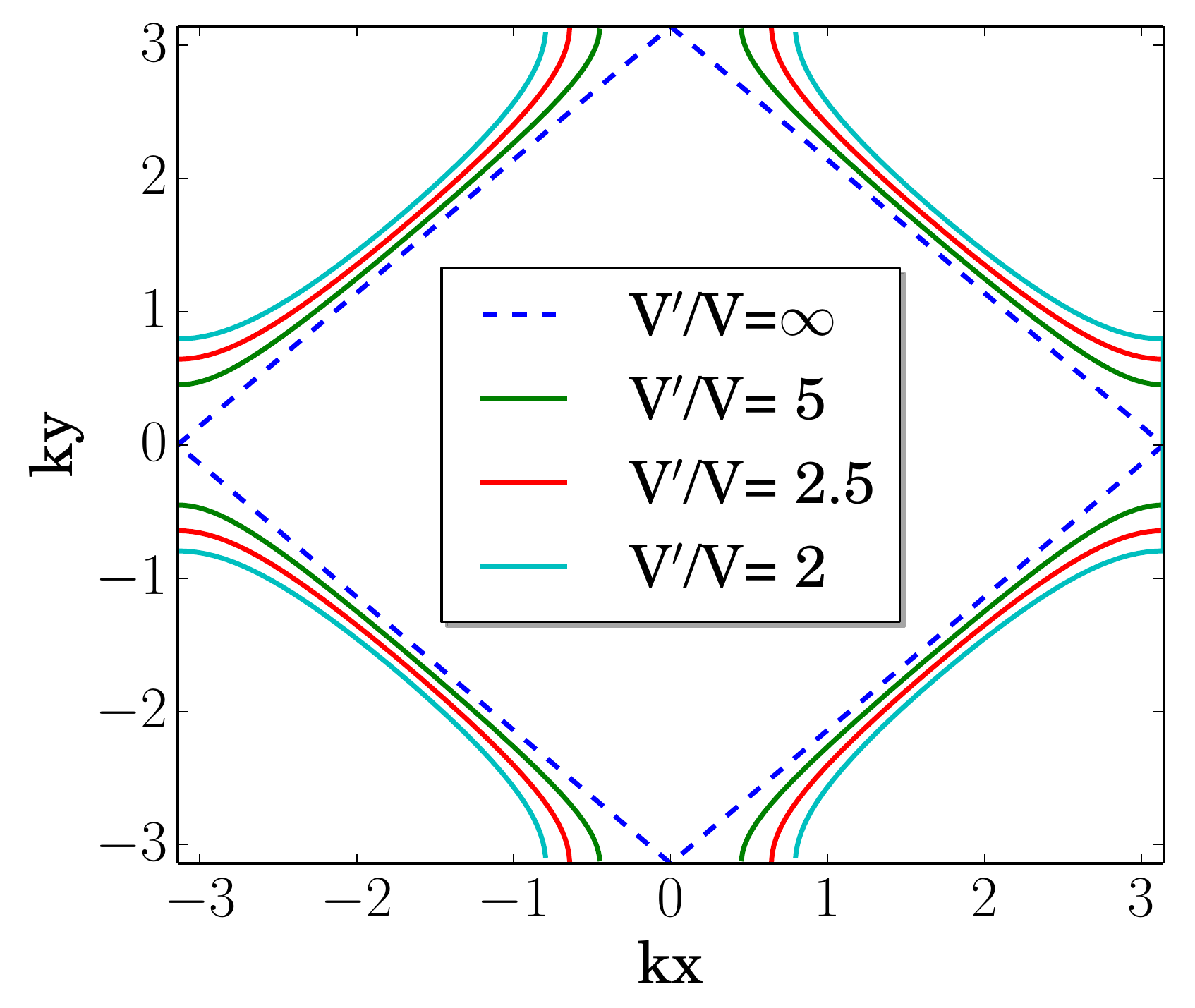}
\caption{(color online) Fermi surfaces of the 2d PAM model for different $V'/V$ ratios.}
\end{figure}

\begin{figure}
\includegraphics[bb=0bp 15bp 240bp 185bp,scale=0.75]{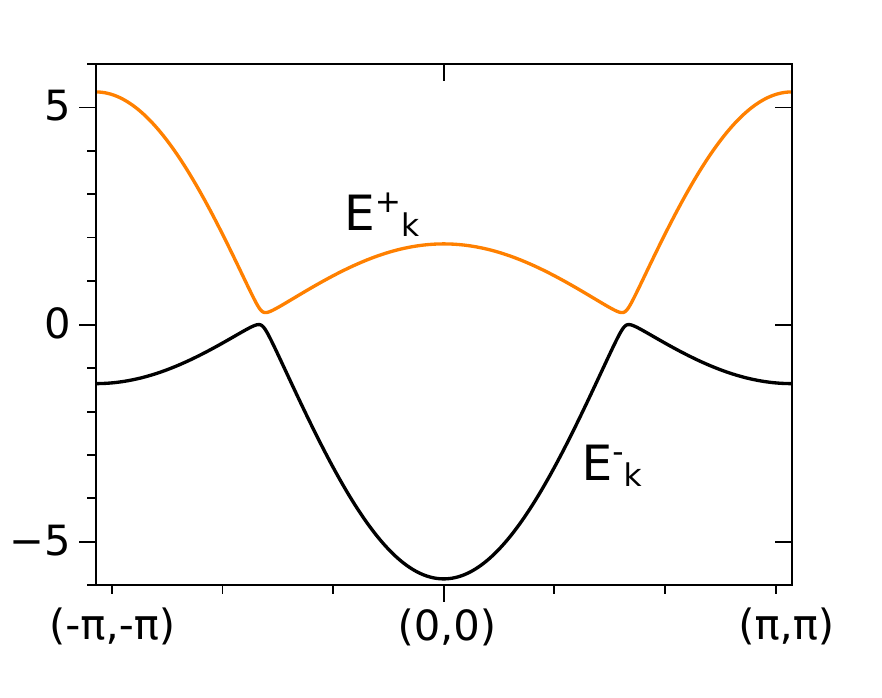}
\caption{(color online) Band structure of the 2d PAM for $V$=0.3 and $V'$=0.75.}
\end{figure}

The Fermi surfaces and band structure in the non-interacting limit are depicted respectively in Fig.~6 and Fig.~7. The Fermi surface resembles the $\alpha$ band of the Ce-115 materials. Pairing occurs mostly in that band in these superconductors.  

The frustrated PAM is a simplified version of the effective Hamiltonian of CeCoIn$_5$ found in Ref.~\onlinecite{allan2013} by fitting quasiparticle interference experiments. That Hamiltonian has 14 parameters. The main lesson we learn from it is that the $f-c$ hybridization is highly frustrated since it is of the same order of magnitude for the on-site, near-neighbor and next-nearest-neighbor terms. Similarly, the two-fluid model in Ref.~\cite{yang_emergent_2012} introduces a RKKY coupling of the form $J_n(\cos(k_x)+\cos(k_y))+J_{nn}\cos(k_x)\cos(k_y)$ which carries the same general effect as our $V'$, namely an angular dependence related to the square-lattice symmetry. 

Note that the hybridization term $V'(\cos(k_x)+\cos(k_y))$  is even under inversion symmetry, like the analogous term in Ref.~\onlinecite{allan2013}.  To understand this symmetry in the Ce$-115$ materials, it suffices to note that the Ce $f$-orbitals, which are odd under parity, couple to the out-of-plane nearest-neighbour $p$-orbitals of Indium, which are also odd.~\cite{Haule2010,shim_modeling_2007}  

\section{DCA method}
Throughout this paper, the DCA calculations are performed for a $2\times2\times2$ cluster, with a weak-coupling continuous
time quantum Monte Carlo (CTQMC) impurity solver.~\cite{Rubtsov2005} In a typical DCA loop, it takes about $5\times 10^7$ CTQMC sweeps to
calculate the Green functions. The Lanczos solver at zero temperature on a $2\times2$ cluster with 7-8 bath sites is used in the section where the real-frequency functions $I_{\Sigma}(\omega)$ and $I_{G}(\omega)$ are computed.
Cellular DMFT (CDMFT) on $2\times2\times2$ cluster has also been carried out to compare with DCA results. Qualitative consistency is obtained, though the superconducting transition $T_c$ in CDMFT is lower than that obtained in DCA. This may be because DCA uses periodic boundary conditions while CDMFT does not.~\cite{ThMaier2005review,KotliarRMP:2006,LTP:2006}

In order to calculate the pairing susceptibilities, we have used the pinning field approach,~\cite{Lin2012} \textit{i.e.}, we observe the response of the system as small pairing fields are applied. To make sure that the response resides in the linear region, we used pinning fields
of three different strengths, $5\times 10^{-5}$, $2\times 10^{-4}$ and $1\times 10^{-3}$ in order to monitor the changes of the pair response function.

\bibliography{heavyfermion2014}


\end{document}